\documentclass[aip,jcp,onecolumn,showpacs,superscriptaddress]{revtex4-1}
\usepackage{graphicx}
\usepackage{amsmath,amssymb}
\usepackage[version=3]{mhchem}
\begin{document}

\title{Surface tensions and surface potentials of acid solutions}

\author{Alexandre P. dos Santos}
\affiliation{Instituto de F\'{\i}sica, Universidade Federal do Rio Grande do Sul, Caixa Postal 15051, CEP 91501-970, 
Porto Alegre, RS, Brazil}
\author{Yan Levin}
\affiliation{Instituto de F\'{\i}sica, Universidade Federal do Rio Grande do Sul, Caixa Postal 15051, CEP 91501-970, 
Porto Alegre, RS, Brazil}

\begin{abstract}
A theory is presented which allows us to quantitatively calculate the excess surface tension of acid solutions. 
The \ce{H+}, in the form of hydronium ion, is found to be strongly adsorbed to the
solution-air interface. 
To account for the electrostatic potential difference measured experimentally, it is necessary to assume that
the hydronium ion is oriented with its hydrogens pointing into the bulk water.  The theory is quantitatively accurate for surface tensions and
is qualitative for electrostatic potential difference across the air-water interface.
 
\end{abstract}

\maketitle
\section{Introduction}

Electrolyte solutions are of fundamental interest for a variety of disciplines.  Over a hundred years ago Hofmeister observed a strong dependence of the  
stability of protein solutions on the specific nature of electrolyte. While some ions tend to stabilize protein solutions, often denaturing them in the process,
others destabilize them favoring protein precipitation. A few years after Hofmeister, Heydweiller~\cite{He10} observed that salt increases
the surface tension of the air-water interface. Furthermore, Heydweiller noticed that the relative effect that ions have on the surface tension
follows closely the Hofmeister series, suggesting that the two phenomena are related.  

Over the last hundred years
there has been a great effort to understand how ionic specificity influences stability of protein solutions and how it affects 
the surface tension of the air-water interface.  
Langmuir~\cite{La17} was probably the first to attempt to construct a quantitative theory of surface tensions of electrolyte solutions.  
Appealing to the Gibbs adsorption isotherm, Langmuir concluded that the excess surface tension of electrolyte solution 
was a consequence of ionic depletion from
the interfacial region. However, no clear explanation for this depletion was provided.  
A few years after, Wagner~\cite{Wa24} argued that this depletion was the result of interaction between the ions and their electrostatic images
across the air-water interface.  Onsager and Samaras~\cite{OnSa34} simplified Wagner's theory and obtained a limiting law which they argued was universally valid for all electrolytes at sufficiently small concentration. 
More recently, Levin and Flores-Mena~\cite{LeMe01}  used a direct free energy calculation to obtain surface tension of strong electrolytes.
To have a quantitative agreement with experiments, these authors stressed the fundamental importance of ionic hydration.  
Nevertheless,  the theory of Levin and  Flores-Mena was not able to predict correctly 
surface tensions of all electrolyte solutions.   Bostr\"om et al.~\cite{BoWi01} 
suggested that Hofmeister effect and the ionic specificity are a consequence of dispersion forces arising from finite 
frequency electromagnetic fluctuations.  This theory predicted that weakly polarizable 
cations should be adsorbed at the air-water interface.  This, however, was contradicted by the experimental measurements of the 
electrostatic potential
difference~\cite{Fr24,JaSc68} and by the
simulations on small water clusters~\cite{PeBe91,DaSm93,StBr99}, as well as by the subsequent large scale polarizable force fields simulations~\cite{JuTo06,JuTo02,HoHe09,Br08} and the photoelectron emission experiments~\cite{MaPo91,Gh05,Ga04}.
These experiments and simulations showed that some anions  --- and not cations --- are present at the solution-air interface~\cite{PeBe91,DaSm93,StBr99,JuTo02,HoHe09,Br08,JuTo06,MaPo91,Gh05,Ga04}. To explain this, Levin~\cite{Le09} extended the traditional Born
theory of ionic solvation to account for ionic polarizability.  The new theory predicted that highly polarizable anions can actually prefer 
an interfacial solvation.  In a followup work, Levin et al.~\cite{LeDo09,DoDi10} used this theory to quantitatively 
calculate the surface tensions and the surface potentials of 10 different
electrolyte solutions and to reproduce the Lyotropic, Hofmeister, series.

While almost all salts lead to increase the air-water surface tension, acids tend to lower it~\cite{WePu96,RaSc66}. The only explanation for this 
is a strong proton adsorption at the water-air interface~\cite{LeHo07,MuFr05,JuTo06,PeIy04,PeSa05,IyDa05}. It is well known that \ce{H+} ion forms various complexes with water molecules~\cite{Ei64,Zu00,MaTu99}.  It has also been suggested that the high surface adsorption of \ce{H+} is related to the hydronium (\ce{H3O+}) geometry~\cite{PeIy04}. This ion, has a trigonal pyramidal structure with the hydrogens located at the base of the pyramid \cite{MuFr05}. In this form hydrogens are good hydrogen-bond donors, while oxygen is a bad hydrogen-bond receptor \cite{PeIy04}.  This favors hydronium ion to be preferentially located at the interface, with the hydrogens pointing towards the bulk water and the oxygen pointing into the gas phase \cite{MuFr05}. Explicit solvation energy calculations confirm this picture \cite{Da03}.

The sign of electrostatic surface potential difference is related with the relative population of cations and anions at the interface. Because of high adsorption of hydronium ions, one would naturally expect that the electrostatic potential difference across the air-water interface 
for acid solutions should be positive. 
The experiments, however, show that the surface potential difference for acids has the same sign as for halide salts, i.e. is predominantly negative \cite{RaSc66,Fr24}. 
Frumkin \cite{Fr24} suggested that this apparently strange behavior might be a consequence of the incomplete dissociation of acid molecules. 
A different explanation was advanced by
Randles \cite{Ra63} who argued that presence of hydroniums at the interface leads to a preferential orientation of water molecules resulting in a dipole layer
with a negative electrostatic potential difference across it.  This conclusion is in agreement with the theory proposed in the present 
paper, as well as with the recent molecular dynamics simulations \cite{MuFr05}.

In this paper we present a theory that allows us to quantitatively calculate surface tensions of acid solutions using only one adjustable parameter
related to the strength of the hydronium adsorption to the interface.  Predictions of the theory are  compared with the experimental measurements. 
The theory is then used to estimate the electrostatic potential difference across the water-air interface for various acid solutions. 

\section{Model and theory}

We consider an acid solution in a form of a drop of radius $R$, where $r=R$ is the position of the Gibbs dividing surface (GDS) \cite{HoTs03,LeDo09}. The water and air will be modeled as uniform dielectrics of permittivities $\epsilon_w=80$ and $\epsilon_o=1$, respectively. The surface tension can be obtained by integrating the Gibbs adsorption isotherm equation:
\begin{equation}\label{ge}
{\rm d} \gamma=-\Gamma_+ {\rm d} \mu_+ - \Gamma_- {\rm d} \mu_- \ ,
\end{equation}
where  $\mu_\pm=k_BT\ln(c_b \Lambda_\pm^3)$ are the chemical potentials and $\Lambda_\pm$ are the de Broglie thermal wavelengths. In this equation $+$ sign
corresponds to the hydronium ion, and $-$ sign to the anion. The bulk ion concentration is $c_b=\rho_+(0)=\rho_-(0)$, where $\rho_{\pm}(r)$ are the ionic density profiles. The ion excess per unit area due to existence of the interface is
\begin{equation}
\Gamma_\pm=\frac{1}{4 \pi R^2} \left[N -\frac{4 \pi R^3}{3} c_b \right] \ ,
\end{equation}
where $N$ is the total number of acid ``molecules''. The ionic density profiles, $\rho_{\pm}(r)$, will be calculated using a modified Poisson-Boltzmann (mPB) equation, as discussed later in the paper.

Anions are divided into two categories: kosmotropes and chaotropes. The theory of electrolyte solutions \cite{DoDi10} showed that chaotropes,  Br$^-$, I$^-$,  NO$_3^-$, and ClO$_4^-$, lose their hydration sheath near the GDS and are partially adsorbed to the interface.   On the other hand 
kosmotropes, F$^-$, Cl$^-$, and SO$_4^{2-}$ remain hydrated in the interfacial region and are repelled from the GDS.

To bring an ion of radius $a_h$ to distance $z>a_h$ from the GDS requires \cite{LeMe01}:
\begin{equation}
W(z;a_h)=\frac{q^2}{2\epsilon_w} \int_0^\infty dk e^{-2 s (z-a_h)} \frac{k[ s \cosh(k a_h)-k \sinh(k a_h)]}{s[ s \cosh(k a_h)+ k \sinh(k a_h)]} \ ,
\end{equation}
of work.  In this equation $s=\sqrt{\left( \kappa^2 + k^2 \right)}$ and $\kappa=\sqrt{8 \pi q^2 c_b/\epsilon_w k_B T}$ is the inverse Debye length.
The kosmotropic ions remain strongly hydrated in the interfacial region and encounter a hardcore-like repulsions from the GDS at a distance of one 
hydrated ionic radius. On the other hand, strongly
polarizable chaotropic anions (Br$^-$, I$^-$, NO$_3^-$ and ClO$_4^-$)  loose their hydration sheath and can move cross the water-air interface. 
However, to avoid the large electrostatic energy penalty of exposing the charge to a low-dielectric (air) environment, 
the electronic charge density of a chaotropic anion redistributes
itself so as to remain largely hydrated \cite{Le09}.  The fraction of ionic charge which remains inside the aqueous environment,  $x(z)$, can be
calculated by minimization the polarization energy \cite{Le09}
\begin{equation}
U_p(z,x)= \frac{q^2}{2 a_0 \epsilon_w}\left[\frac{\pi x(z)^2}{\theta(z)}+\frac{\pi [1-x(z)]^2 \epsilon_w}{[\pi-\theta(z)]\epsilon_o}\right] + \\
\frac{(1-\alpha)}{\alpha \beta} \left[ x(z)-\frac{1-cos[\theta(z)]}{2} \right]^2 \ .
\end{equation}
In the above equation $\alpha$ is the relative polarizability defined as $\alpha=\gamma_i/a_0^3$, where $\gamma_i$ is the ionic polarizability, $a_0$ is the unhydrated (bare) radius, and $\theta(z)=\arccos[-z/a_0]$. Performing the minimization we obtain
\begin{equation}
x(z)=\left[ \frac{\lambda_B \pi \epsilon_w}{a_0 \epsilon_o \left[\pi-\theta(z)\right]}+\frac{(1-\alpha)}{\alpha} [1-cos[\theta(z)]] \right] / \left[\frac{\lambda_B \pi}{a_0 \theta(z)} + \frac{\lambda_B \pi \epsilon_w}{a_0 \epsilon_o [\pi-\theta(z)]} +2\frac{(1-\alpha)}{\alpha} \right] \ ,
\end{equation}
where $\lambda_B=q^2/\epsilon_w k_BT$ is the Bjerrum length.

\begin{figure}[t]
\begin{center}
\includegraphics[width=7cm]{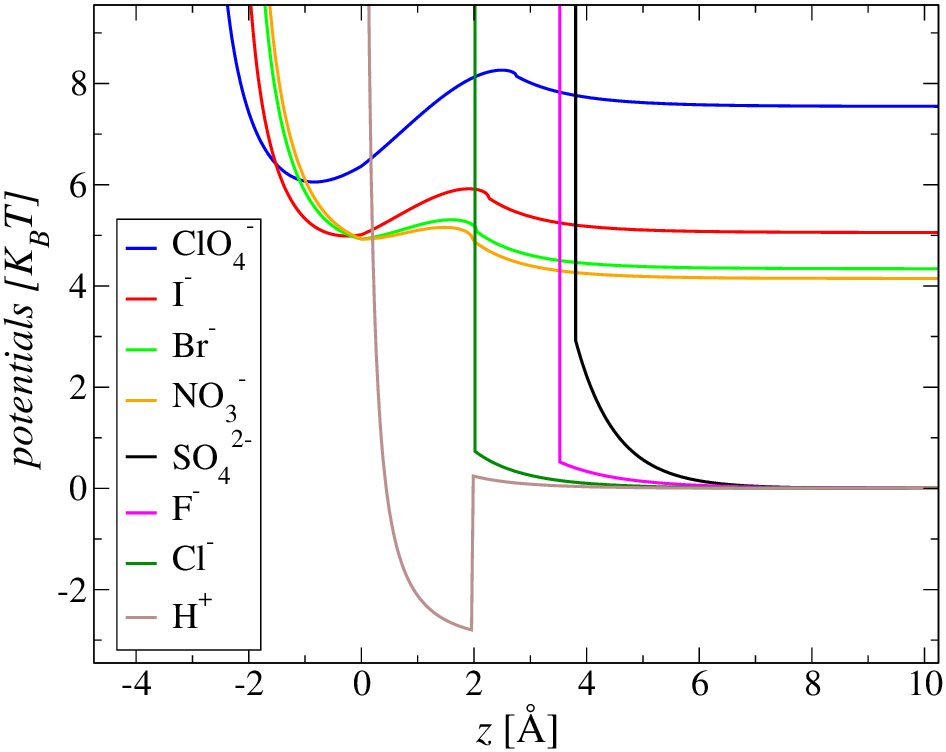}
\end{center}
\caption{Potentials for all ions at $1$M. The GDS is at $z=0$ $\,$\AA. The bulk cavitational potential is not considered for kosmotropes, since it does not change along the drop.}
\label{fig1}
\end{figure}

The force that drives chaotropic ions towards the interface results from water cavitation. 
To introduce an ion into an aqueous environment requires creating a cavity, which perturbes the hydrogen bond network of water molecules. 
For small ions, the free energy cost of 
forming a cavity is entropic and is proportional to the volume of the void formed \cite{LuCh99,Ch05}. As the ion moves across the
GDS, its cavitational free  energy decreases.  This results in a short-range attractive potential between the anion and the GDS \cite{LeDo09}:
\begin{eqnarray}\label{cavpot}
U_{cav}(z)=\left\{
\begin{array}{l}
 \nu a_0^3 \text{ for } z \ge  a_0  \ , \\
 \frac{1}{4} \nu a_0^3  \left(\frac{z}{a_0}+1\right)^2 \left(2-\frac{z}{a_0}\right) \text{ for } -a_0<z<a_0 \ ,
\end{array}
\right.
\end{eqnarray}
where $\nu \approx 0.3 k_B T/$ \AA$^3$ is obtained from bulk simulations \cite{RaTr05}.  For hard (weakly polarizable ions) the cavitational free energy
gain is completely overwhelmed by the electrostatic free energy penalty of moving ionic charge into the low dielectric environment.  For soft polarizable
ions, however, the electrostatic penalty is small, since most of the ionic charge remain inside the aqueous environment.  The total potential of a soft anion, therefore, has a minimum in the vicinity of the GDS, see Fig. \ref{fig1}.

The H$^+$ ions (protons) do not exist as a separate specie in water. Instead they form complexes with water molecules, H$_3$O$^+$ and H$_2$O$_5^+$ \cite{Ei64,Zu00,MaTu99}. Because of its favorable geometry (trigonal pyramidal), the hydronium ion (H$_3$O$^+$) adsorbs to the water-air interface with a preferential orientation \cite{MuFr05} of oxygen towards the air.  We model this attraction by a square well potential with a range of a 
hydrogen bond  $1.97$ \AA,
\begin{eqnarray}\label{uhtotal}
U_{hyd}(z)=\left\{
\begin{array}{l}
0 \, \text{ for } z  \ge  1.97\ \text{\AA} \ , \\
-3.05 \ k_BT \, \text{ for } 0 \le z < 1.97\ \text{\AA} \ .
\end{array}
\right.
\end{eqnarray}
The depth of the potential is then adjusted to obtain the experimentally measured surface tension of HCl.  The same potential is then used to calculate
the surface tensions of all other acids.  We should stress, however, that one should
not attach too much meaning to the specific value of the potential depth.  The real proton transfer is a 
quantum mechanical process, therefore there is bound to be some arbitrariness in how one models it at a classical level.  Here we have chosen the 
range of the square well potential to be one hydrogen bond.  If one changes this distance, the depth of the potential will
have to be modified to obtain an optimal fit of the surface tension of HCl solution. 
However once this is done, the values of the surface tension of other acids will not be significantly affected. 
Thus, the strength of   H$^+$  potential 
is the only free parameter of the theory. The total potential felt by  H$^+$ is then,  $U_H(z)=U_{hyd}(z)+W(z;0)$, see Fig. \ref{fig1}.

While the kosmotropic anions feel only the potential $W(z;a_h)$ and the hardcore repulsion from the GDS, the chaotropic anions are influenced by the total potential \cite{LeDo09}:
\begin{eqnarray}
U_{tot}(z)=\left\{
\begin{array}{l}
W(z;a_0)+\nu a_0^3+\frac{q^2}{2 \epsilon_w a_0} \, \text{ for } z \ge  a_0 \ , \\
W(a_0;a_0) z/a_0 + U_p(z)+U_{cav}(z)\, \text{ for } 0<z<a_0 \ , \\
U_p(z)+U_{cav}(z)\, \text{ for } -a_0<z \le 0 \ .
\end{array}
\right.
\end{eqnarray}
In Fig. \ref{fig1}, we  plot the potentials felt by various ions at $1$M concentration, as a function of the distance from the GDS.

\begin{table}[t]
\centering
\setlength{\belowcaptionskip}{10pt}
\caption{Ion classification into chaotropes (c) and  kosmotropes (k). Effective radii (hydrated or partially hydrated) for kosmotropes and (bare) for chaotropes, for which we have also include the polarizabilities from Ref. \cite{PyPi92}.}
      \begin{tabular}{|c|c|c|c|}
      \hline
      Ions          &    chao/kosmo     & radius (\AA) &  polarizability (\AA$^3$) \\
      \hline
      F$^-$         &        k          &     3.54     &            *              \\
      \hline
      Cl$^-$        &        k          &     2        &            *              \\
      \hline
      Br$^-$        &        c          &     2.05     &           5.07            \\
      \hline
      I$^-$         &        c          &     2.26     &           7.4             \\
      \hline
      NO$_3^-$      &        c          &     1.98     &           4.48            \\
      \hline
      ClO$_4^-$     &        c          &     2.83     &           5.45            \\
      \hline
      SO$_4^{2-}$   &        k          &     3.79     &            *              \\
      \hline
   \end{tabular}
   \label{tab1}
\end{table}
The ionic density profiles can now be obtained by integrating the mPB equation:
\begin{eqnarray}
\nabla^2 \phi(r)&=&-\frac{4\pi q }{\epsilon_w} \left[\rho_+(r)-\rho_-(r)\right] \ , \\
\rho_+(r)&=&\frac {N e^{-\beta q \phi(r)-\beta U_H(z)}}{\int_0^{R} 4\pi r^2\, dr\, e^{-\beta q \phi(r)-\beta U_H(z)}} \nonumber \ , \\
\rho_-^{chao}(r)&=&\frac{N e^{\beta q \phi(r) -\beta U_{tot}(r)}}
{\int_0^{R+a_0} 4\pi r^2\, dr \, e^{\beta q \phi(r) -\beta U_{tot}(r)}} \nonumber \ , \\
\rho_-^{kos}(r)&=&\frac {N \Theta (R-a_h-r) e^{\beta q \phi(r)-\beta W(z;a_h)}}{\int_0^{R-a_h} 4\pi r^2\, dr\, e^{\beta q \phi(r)-\beta W(z;a_h)}} \nonumber \ ,
\end{eqnarray}
where $\Theta$ is the Heaviside step function, $\rho_-^{chao}(r)$ is the density profile for chaotropic anions and $\rho_-^{kos}(r)$ for kosmotropic ones.

Once the ionic density profiles are calculated, the surface tensions can be obtained by integrating the Gibbs adsorption isotherm (eq \ref{ge}). The ionic radii and polarizabilities are the same as were used in our previous work on surface tension of electrolyte solutions \cite{LeDo09,DoDi10}. In Table \ref{tab1} we summarize this data.

\begin{figure}[t]
\begin{center}
\includegraphics[width=7cm]{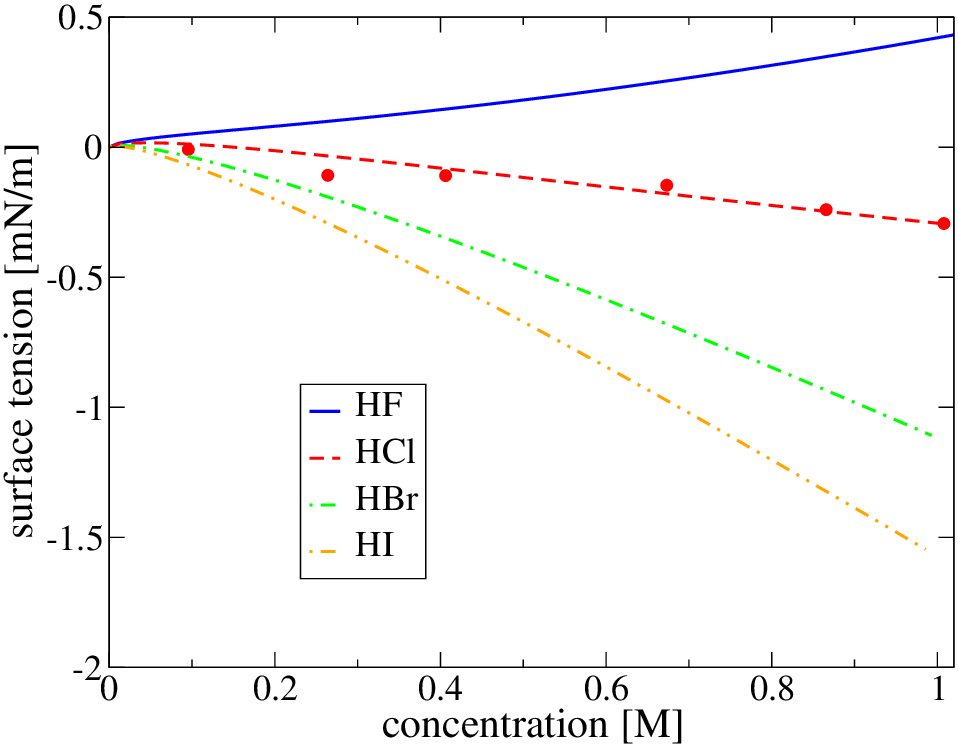}
\end{center}
\caption{Surface tensions for HF, HCl, HBr, and HI. The symbols are the experimental data for HCl \cite{WePu96} and the lines are the results of the present theory.}
\label{fig2}
\end{figure}

\begin{figure}[h]
\begin{center}
\includegraphics[width=7cm]{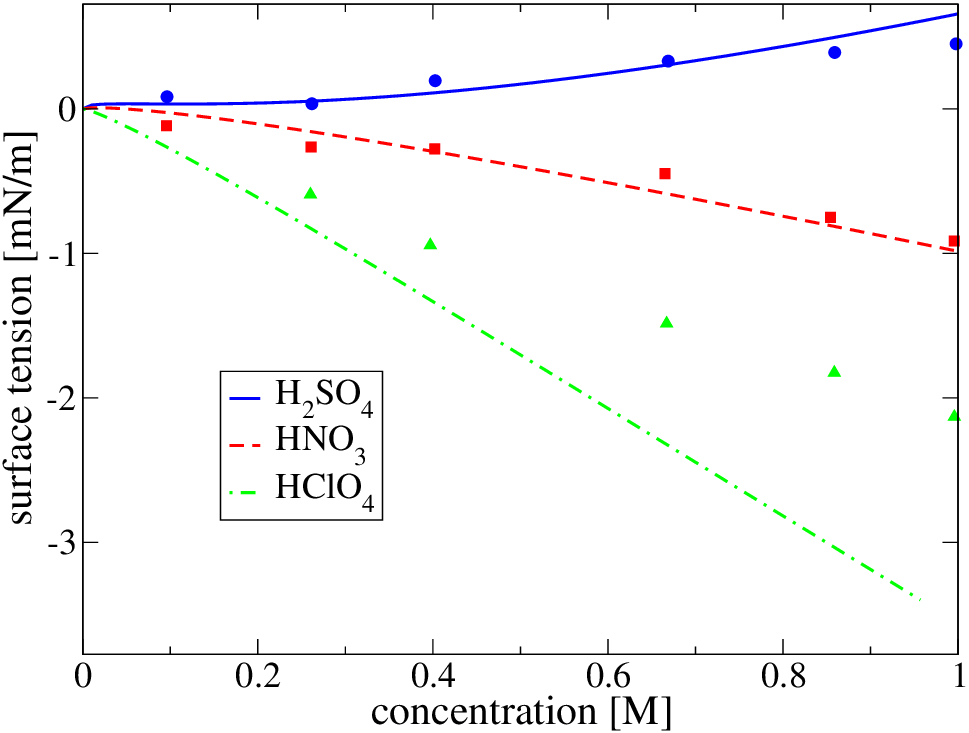}
\end{center}
\caption{Surface tensions for H$_2$SO$_4$, HNO$_3$, and HClO$_4$. The symbols are the experimental data \cite{WePu96} and the lines are the results of the present theory.}
\label{fig3}
\end{figure}

The depth of the potential $U_{hyd}(z)$ (eq \ref{uhtotal}) is adjusted 
to fit the HCl experimental data \cite{WePu96} (see Fig. \ref{fig2}), this is the only adjustable parameter of the theory.  
We find that a square well potential of depth $-3.05$ $k_BT$  
results in an excellent fit of the experimental data for HCl  in the range of concentrations from $0$ to $1$ M.
The excess surface tension of  all other acids is then calculated using the same potential $U_{hyd}(z)$. 
The predictions for the surface tensions of HF, HBr, and HI are plotted in Fig. \ref{fig2}.  Unfortunately, we have no experimental data
to compare for these halogen acids. For H$_2$SO$_4$ and HNO$_3$ (see Fig. \ref{fig3}),  we find a good agreement between the theory and experiment. For HClO$_4$ the theory overestimates the surface tension. This is similar to what was found for sodium perchlorate salt \cite{DoDi10}.  The difficulty is that ClO$_4^-$ is a large weakly hydrated ion.  Since the cavitational energy grows with the cube of ionic radius, a small error in radius leads to a big error in surface tension.


Finally, we use the theory to calculate the electrostatic potential difference across the solution-air interface.
The surface potential difference $\Delta \chi=\phi(R+a_0)-\phi(0)$ predicted by the present theory has a wrong sign compared to Frumkin's experimental measurements --- positive instead of negative \cite{RaSc66,Fr24}.  Positive sign reflects a strong adsorption of hydronium ions to the GDS. The simple 
dielectric continuum theory presented here, however, does not account for the structure of the interfacial water layer.  Since the
hydronium ion at the GDS has a preferential orientation with the hydrogens pointing towards the bulk,  presence of many such ions will
result in a dipole layer.  Note that in the absence of hydroniums, the water dipoles predominantly point along the interface \cite{KaKu08}.
The hydronium layer produces an electric field $E=4 \pi p N_h/\epsilon_o d A$, where $N_h$ is the number of hydroniums at the interface,
$p$ is the water dipole moment, $d$ is the dipole length, and $A$ is the interfacial area.  If we suppose that all the hydroniums are perfectly aligned, the potential difference across the dipolar layer will be  $\Delta \chi_w=- 4 \pi p \Gamma_+/\epsilon_o$. Using the dipole moment of a
water molecule, $p=1.85$ D, we obtain the dipole layer contribution to the overall electrostatic potential difference. 
Adding this to $\Delta \chi$, we obtain the total electrostatic surface potential difference across the solution-air interface. 
In Table \ref{tab2}, we list the surface potentials of various acids at 1M concentration.
\begin{table}[t]
\centering
\setlength{\belowcaptionskip}{10pt}
\caption{Surface potential differences for various acids at 1M concentration. Contributions from electrolyte and aligned water dipoles.}
      \begin{tabular}{|c|c|c|}
      \hline
      Acids         &  calculated [mV]   & Frumkin \cite{Fr24} [mV] \\
      \hline
      HF            &        $85.5$      &         $-71$             \\
      \hline
      HCl           &        $1.24$      &         $-23$             \\
      \hline
      HBr           &        $-95$       &         $-34$             \\
      \hline
      HI            &        $-144.8$    &         $-61$             \\
      \hline
      HNO$_3$       &        $-84.4$     &         $-48$             \\
      \hline
      HClO$_4$      &        $-412$      &         $-82$             \\
      \hline
   \end{tabular}
   \label{tab2}
\end{table}
Clearly these values are an exaggeration of the total electrostatic potential difference across the interface, since at finite temperature
there will not be perfect alignment of interfacial hydronium ions.
Nevertheless,  the theory should provide us an order of magnitude estimate of the electrostatic potential difference.  
In fact, for most acids we find a reasonable agreement between the predictions of the theory and Frumkin's experimental measurements~\cite{Fr24}.  
A noticeable exception is the HF. Experimental potential for hydrogen fluoride measured by Frumkin is negative, 
while we find a large positive value.  Frumkin's value for HF is clearly
outside the general trend for halogen acids.  In his classical review of electrolyte solutions, Randles \cite{Ra63} did not mention Frumkin's result
for HF acid, while discussing his other measurements.  We can only suppose that Randles also did not have a complete 
confidence in this particular value.
Experimental measurements of excess surface potentials are very difficult.  
This is probably the reason why Frumkin's measurements of surface
potentials of acids have not been repeated in over 90 years. 

\section{Conclusions}
In this paper we have developed a theory for surface tensions of acid solutions. The hydronium adsorption to the interface was modeled by a square well potential, the depth of which is the only adjustable parameter of the theory.  The agreement between the theory and experiments is very reasonable
for different acid solutions at concentrations from 0 to 1M.   In order to account for the experimental values of the excess electrostatic surface potential, we must require a preferential orientation of hydronium ion at the interface, with the hydrogens pointing into the bulk.  With this assumption we get 
a qualitative agreement with the experimental measurements of the excess electrostatic potentials of various acid solutions. At the moment this is
the only theory that can account (quantitatively)  for the surface tensions and (qualitatively) for the surface potentials of acid solutions. 
\section{Acknowledgments}
This work was partially supported by the CNPq, INCT-FCx, and by the US-AFOSR under the grant FA9550-09-1-0283.

\end{document}